\documentclass{amsart}
\usepackage{graphicx}
\usepackage{epsf}
\theoremstyle{definition}

\theoremstyle{remark}

\numberwithin{equation}{section}

\newcommand{\bt}{\begin{tabular}}
\newcommand{\et}{\end{tabular}}
\newcommand{\bi}{\begin{itemize}}
\newcommand{\ei}{\end{itemize}}
\newcommand{\bbm}{\begin{minipage}[h]{6cm} ${}$ \\}
\newcommand{\eem}{\\ ${}$ \end{minipage}}
\newcommand{\bmm}{\begin{minipage}[h]{1.2truecm} ${}$ \\}

\begin{document}

\vspace{-4truecm}
{}\hfill{DSF$-$6/2005}
\vspace{1truecm}

\title{Ettore Majorana's course on Theoretical Physics: a recent discovery}%
\author{A. Drago}%
\address{{\it A. Drago}: Gruppo di Storia della Fisica, Dipartimento di Scienze Fisiche,
Universit\`a di Napoli ``Federico II'', Complesso Universitario di M. S. Angelo, Via Cinthia, 80126 Napoli}%
\author{S. Esposito}%
\address{{\it S. Esposito}: Dipartimento di Scienze Fisiche,
Universit\`a di Napoli ``Federico II'' \& I.N.F.N. Sezione di
Napoli, Complesso Universitario di M. S. Angelo, Via Cinthia,
80126 Napoli ({\rm Salvatore.Esposito@na.infn.it})}%


\begin{abstract}
We analyze in some detail the course of Theoretical Physics held by Ettore Majorana at the University of Naples in 1938, just before his mysterious disappearance. In particular we present the recently discovered "Moreno Paper", where all the lecture notes are reported. Six of these lectures are not present in the collection of the original manuscripts conserved at the Domus Galilaeana in Pisa, consisting of only ten lectures. 
\end{abstract}
\maketitle

\section{Introduction}

\noindent Probably, the highest appraisal received by the work of Ettore Majorana was expressed by Enrico Fermi writing to another physicist, Giuseppe Cocconi. Just after the mysterious disappearance of Majorana in the March of 1938, in order to explain to Cocconi what this loss did really mean for the physicists' community, Fermi expressed himself as follows:
``... but then there are geniuses like Galileo and Newton. Well, Ettore was one of them. Majorana had what no one in the world has..." \cite{Cocconi}. Similar opinions were expressed by Fermi in several other, more formal in nature, situations as, for instance, in formalizing the judgement given by the board of the 1937 competition for a full professorship in Theoretical Physics (a position which was obtained by Majorana independently from the common competition rules) or in a letter to the Italian Prime Minister, Mussolini, where Fermi urged him for performing accurate searches about Majorana disappearance \cite{Recami}.
Such opinions could appear as overstatements or unjustified (especially because they are expressed by a great physicist as Fermi), when compared with the spare (known) Majorana's scientific production, just 9 published papers; however, today the name of Majorana is largely known to the nuclear and subnuclear physicist's community (Majorana neutrino, Majorana-Heisenberg exchange forces, and so on).
\\
Some light on the peculiar abilities of Majorana is brought to us by his wide unpublished scientific papers, almost all deposited at the Domus Galilaeana in Pisa; those known, in Italian, as ``Volumetti" has been recently collected and translated in a book \cite{volumetti}. 
\\
In this paper we focus our attention on Majorana as professor of Theoretical Physics rather than as researcher, and found our analysis mainly on the remembrances of his former students, as well as on the lecture notes for the course given at the University of Naples. We refer especially to the important recent discovery of handwritten lecture notes containing, in particular, six previously unknown lectures beyond the ten ones whose original manuscripts are conserved in Pisa. Apart from the scientific content of these documents, the historical relevance of such discovery is given by the detailed chronological registration of the development of the course and even of the last days of Majorana in Naples before his disappearance, which will be established here by making use also of some other evidences, more or less known, collected on Majorana as a teacher. For a better and easier understanding, before such an analysis we quickly review the main events which led Majorana to teach at the University of excellent biographies \cite{Recami}, \cite{Amaldi}.

\section{The national professor competition and the arrival in Naples}

\noindent In 1937 the University of Palermo (Sicily), upon request of Emilio Segr\'e, called for a novel competition for a full professorship in Theoretical Physics. In addition to Majorana (which was urged to take part to the competition by his friends
and Fermi), the competitors were Leo Pincherle, Giulio Racah, Gleb Wataghin, Gian Carlo Wick and Giovanni Gentile (the son of the homonymous philosopher and former Ministry of Education). The chairman of the competition board, installed in Rome, was Fermi himself, while the other members were Antonio Carrelli (secretary), Orazio Lazzarino, Enrico Persico and Giovanni Polvani.
\\
The official documents certify that the competition board suggested to the Ministry of Education G. Bottai (which then accepted the proposal) ``to appoint Majorana as full professor of Theoretical Physics in a University of the Italian kingdom, for high and well-deserved repute, independently of the competition rules" \cite{concorso}. Once the chair was attributed to Majorana ``out of competition", the board then declared the three winning competitors according to the following order: 
1) G.C. Wick, 2) G. Racah, 3) G. Gentile.
Notice that one member of the competition board, A. Carrelli, at that time was the Director of the Institute of Physics at the University of Naples, and probably he played some role in suggesting this place as the affiliation for Majorana's chair.
In fact, Carrelli (mainly an experimentalist) was well aware of the fact that the Institute of Physics in Naples was lacking, at that time, of theoreticians working on frontier physics topics; moreover, from a letter from Majorana to Gentile \cite{Gentile} we learn that Majorana was ``in correspondence with Carrelli, which is really a very good person" 
\footnote{It is also relevant that, before his appointment as Director of the Institute of Physics in Naples, Carrelli directed for some time the Institute of Physics in Catania, the hometown of Majorana.}.
\\
Majorana was informed of his designation as full professor by the Minister Bottai on November 2, 1937. However, his appointment officially started on the 10th of the same month, but he went to Naples not before the beginnings of the following year (probably January 10). Here Majorana realized very soon that very few physicists worked at the Institute of Naples: ``...Practically, the Institute is composed only by Carrelli, his old-dated first assistant [{\it aiuto}] Maione and of the young assistant Cennamo. There is also a professor in Earth Physics, which is, however, difficult to be founded..." \footnote{Alfredo Maione obtained his Degree in Physics in 1933 with Carrelli as advisor; he was appointed as assistant in the same year and became first assistant in 1937. Francesco Cennamo, instead, was assistant to Carrelli from 1934-5 to 1940-1 (he became first assistant in 1940 when Maione resigned from his position). The ``elusive" professor in Earth Physics was probably the Director of the Institute of Earth Physics at the University of Naples, Giuseppe Imb\'o, which however did not belong to the Institute directed by Carrelli.} \cite{MFN1}. Also some students of his course \cite{Sciuti}, \cite{Senatore} testify the fact that Majorana was initially puzzled by the situation found in Naples (even the number of students in Physics was very low, just five as we will see below), but later he probably changed of mind and, in a letter to Gentile \cite{MGN1}, he stated to be ``pleased by my students, some of which appear to have taken Physics seriously".

\section{The course on Theoretical Physics}

\noindent Before the arrival of Majorana in Naples, the course on Theoretical Physics (a third- and fourth-year course) was usually delivered by the Director Carrelli, and the topics covered in it did not concern at all the modern developments of Quantum Physics. Just as an example, Sebastiano Sciuti recalls ironically that one of the most {\it advanced} argument treated in the course by Carrelli was the brownian motion... \cite{Sciuti}.

\subsection{The opening lecture}

\noindent Majorana announced the opening of his course to be held in January 13 (Thursday) at 9 a.m., but he agreed with the academic dean ``to avoid any official character at the opening of the course" \cite{MFN1}. An indirect track of this fact is the absence of such a piece of news on the city newspapers (such as, for example, ``Il Mattino"), differently from what occurred for others (certainly most crowded) academic courses.
\\
According to his student Gilda Senatore \cite{Senatore}, the students could not take part to the opening lecture, commonly considered as a ``lectio magistralis", due to an explicit request by the Director of the Institute or, maybe, to other practical reasons (as, for example, the concomitance of other courses). Majorana himself, in fact, wrote that ``it has been not possible to check that no overlap [with other courses] occurs, so that it is possible that the students do not come and the opening lecture will be postponed'' \cite{MFN1}. As a matter of fact, the opening lecture was given on January 13, as scheduled, in the great semi-cyclical classroom in the building in Via Tari 3 (where the course of Experimental Physics was usually delivered), without students and with the presence of about ten people, according to what Sciuti remembers \cite{Sciuti}. The exclusion of the students from the opening lecture by Carrelli (or some other people) may be convincingly explained by an ancient custom of the University of Naples, according to which the novel teacher should ``demonstrate" to the professors of the Neapolitan athenaeum to be deserving of the assigned chair \cite{Preziosi}. However, Majorana's quote ``to avoid any official character" reported above and the presence at the opening lecture of some Majorana's relatives \cite{Recami}, would force to think at this interpretation of the exclusion of the students as only one possible guess.
\\
The following lecture, the real start of the course, was given on January 15 (Saturday), and the course continued until March, in the even days of the week (Tuesday, Thursday and Saturday), in a small room put in front of that used for the opening lecture \cite{Sciuti}.

\subsection{The students of the course}

\noindent The ``Physics" students who took part to the Majorana course were five: Filomena Altieri, Laura
Mercogliano, Nada Minghetti, Gilda Senatore and Sebastiano Sciuti. All the women were ``internal students" ({\it allievi interni}), so that while attending the academic courses, they performed also some research activities (mainly in Classical Physics) in parallel with; at this time they were fourth-year students \cite{Senatore}, probably out-of-course \footnote{Note that in the 1930's, the course on Theoretical Physics consisted of two parts (as well as some other courses), delivered at the third and fourth year in Physics, respectively. With the exception of G. Senatore, which defended her Master Thesis one year later, the remaining four students graduated in Physics in December 1938.} \cite{Sciuti}. Sciuti, an experimentalist himself, was preparing his Degree in Physics, and he already had attended a course on Theoretical Physics given by Carrelli. His purpose was to establish some contact with the group of Rome directed by Fermi (he entered in Physics when he was 17th years old, just accepting the general invitation by Orso Mario Corbino and Fermi to study Physics): to attend a course given by a member of the group of Rome appeared to him as a first step in pursuing his aim. In their research activities, all the students were working mainly in the field of Experimental Physics \footnote{The specific topics of their activities can be deduced from the titles of their Master Theses: {\it On the spectral emission at ultrared frequencies of some phosphates and silicates} (Altieri), {\it Infrared absorption bands due to the presence of the OH group} (Mercogliano), {\it
Total metal emission below and above the melting point} (Minghetti), {\it Absorption and fluorescence of quinine salts} (Senatore), {\it Preliminary researches at ultrared frequencies} (Sciuti).}; for example Senatore, although willing to study Theoretical Physics, attended some studies on Molecular Physics being Maione (and, after the disappearance of Majorana, Cennamo) as supervisor \cite{Senatore}.
\\
In addition to the five Physics students mentioned above, at the Majorana course we find also some non-examination students: Mario Cutolo \cite{Senatore}, \cite{Sciuti} and Father Savino Coronato \cite{Senatore}. According to Senatore, Cutolo probably came to this course mainly because of the presence of a girl (Minghetti). Coronato was a student in Mathematics (he graduated at the end of 1938 with Renato Caccioppoli as supervisor) and became the right hand of Caccioppoli at the Institute of Calculus; likely, he was invited by Caccioppoli himself (who was presumably present at the opening lecture) to attend Majorana's lessons. 
\\
In addition to the above information, already known, we have recently realized \cite{Moreno} that one more non-examination student probably participated at the Majorana course: Eugenio Moreno. Such a presence was, until now, neither suspected nor supported by documents, since the living witnesses of the course (mainly Sciuti and Senatore) did not mention him. The new information has, instead, recently emerged due to the son Cesare of E. Moreno (dead in 2000) \cite{Moreno}, and is apparently proved by the exhibition  of the lecture notes handwritten by E. Moreno after the original manuscripts by Majorana.
\\
E. Moreno was born in Naples on February 16, 1910 and entered in Mathematics in the academic year 1929-1930. Due to some military employments, he obtained his Degree in Mathematics relatively late, only on the June 17, 1940. Soon after December 30, 1937, Moreno came back from his call-up service and probably matured the intention to attend Majorana's course on Theoretical Physics which, differently from other courses of the Faculty of Sciences whose openings were usually held on the starting week of the academic year in November, began just in the next January 1938.

\subsection{Majorana's personal style}

\noindent Majorana, ``dressed in blue" \cite{Sciuti}, had always a ``sad and perplexed" look so that his friends of the Fermi's group in Rome since long time attributed to him the nickname ``Great Inquisitor" \cite{Amaldi}, \cite{Recami}. This fact, in addition to the hard subjects covered in his course \footnote{Senatore remembers that the Mathematics used by Majorana was not at all introduced in the other academic courses given at that time.}, certainly embarrassed the young students. Outside the ``official" framework of the course, Majorana confirmed as well his typical behavior: sometimes ``bashful smiled; we guessed that he should be a very good and kind man, but never was affectionate and inviting; he was always extremely shy" \cite{Sena98}. Still: ``in that long and dark corridor at the ground floor... he walked always very near the wall, noiseless and alone, moving as a shadow" \cite{Sena98}.
\\
When he arrived in Naples, certainly the Director Carrelli talked with Majorana about the students and their research activities, so that Majorana probably soon realized of his quite peculiar task to teach the modern Theoretical Physics to such a low number of low-level students \cite{Senatore}. Nevertheless, he was firm in pursuing his task in a very responsible way \cite{Sena98}.
\\
When teaching, ``he was very clear in treating the argument proposed by him at the beginning of the lesson and he carried out it by means of a lot of examples, always emphasizing the physical content rather than the mathematical formalism; but when he turned around the blackboard and succeeded to write, he performed calculations whose meaning were not immediately easy to follow'' \cite{Sena98}. The peculiar character of Majorana, then, certainly did not invite the mild student to stop him for asking some more words of explanation. Few times, some questions were posed by Sciuti only \cite{Sciuti}, who also asked Majorana a textbook in order to follow his lectures in an easier way. To this request, Majorana answered that he would have distributed some notes and, in any case, he was following Persico's textbook \cite{Persico} (``a very fine book, in Italian''), although adding some ``formal simplifications'' \cite{Sciuti}. However, on the basis of the lecture notes came to us (see the next section), we see that such ``simplifications'' were not at all occasional and of a minor role, so that the setting-out given by Majorana to his course was completely distinct from that used by Persico \cite{Drago}. Another book ``suggested'' to Sciuti was Heisenberg's \cite{Heisenberg}, but similar considerations apply to it too \footnote{We point out that both the texts cited above belonged to the little (just about 15 books) personal library of Majorana; in particular, the Heisenberg's text was owned in its original German version of 1930 \cite{EMJR}. Moreover, the reference to the text by Persico seems to appear on the back cover of the Moreno Paper, although partially erased (it was pencilled).} Apparently, then, Majorana suggested an introductory book to a subject which he dealt with in an advanced way.
\\
Thus, when a given topic resulted too hard to be understood, the students could not have recourse to a textbook, and Senatore recalls that ``our notes only, taken during the lessons and compared one with another after the course, allowed us to relate the theoretical part, illustrated in an excellent way, with the mathematics representing it'' \cite{Sena98}. In fact, the students of the course (probably ``Physics'' students only) usually met together in the day after a given lesson in order to ``compare'' the notes taken during the lesson and study the corresponding arguments. Sometimes during a lessons, Majorana noticed (by stopping himself and turning back) that the students had difficulties to understand what he was expounding; then he stopped himself and explained again the same argument \cite{Senatore}. ``Exactly during one of that unpleasant and hard lessons, where the topic was essentially some mathematical tool to be applied to the study of given physical phenomena, Majorana forgot what a high level scientist he was since, when he stayed at the blackboard and wrote on it, suddenly stopped, then turned back, looked us for a while, smiled and presented again the explanation, but by making some effort in relating the concept already exposed with the formulae filling the blackboard'' \cite{Sena98}.

\section{The University lectures}

\noindent As seen above, according to what Sciuti remembers \cite{Sciuti}, Majorana prepared some lecture notes to be given to his students in order to facilitate their understanding of the arguments developed in the course. Probably (see below) this took place after January 22, 1938, i.e. after his fifth lesson, since all the documents currently known are lacking just of the first five lectures (or, more precisely, the opening one and the following four lectures). The story of how such notes, which are particularly important in order to understand the innovative work of Majorana as a teacher, have reached us is rather intriguing, so that it could be worth-while to analyze it with the help of all documents in our hands.

\subsection{The discovery of Moreno Paper}

\noindent On September 2004, the present authors have gathered the valuable report of the Eugenio Moreno's son Cesare, according to which his father would have been one of the most diligent non-examination students of the Majorana course. In this occasion he showed to us an important document (for the sake of brevity, we will call it Moreno Paper) where E. Moreno faithfully copied Majorana's lecture notes.
\\
Without this document, the interested people had, at their own disposal, only few handwritten manuscripts by Majorana, deposited at the Domus Galilaeana in Pisa (Italy), and published some years ago \cite{Bibliopolis}. An analysis of such manuscripts led to the conclusion that, except at most for one or two lectures, these notes were believed to cover substantially the whole Majorana's course on Theoretical Physics \cite{Cabibbo}. Some doubts (mainly based on too far reminds) were risen just by a girl-student of the course, G. Senatore, who always declared that
``some chapters of the lectures are lacking; their text (completed by the notes of the lecture the professor would have delivered in the day subsequent his disappearance) was given to me. Few other sheets lacks yet, who were written in original and in a neat way as the other ones, but they did not correspond to the lectures already given'' \cite{Sena98}.
\\
The relevant importance of the Moreno Paper is twofold: on one hand, it is arrived to us through a completely different path from that followed by the original manuscripts (see below); on the other hand, in the Moreno Paper {\it all} the lectures of the course are {\it numbered} and {\it dated}, differently from Pisa manuscripts (where some lectures (the first ones) are dated, while the others (from that corresponding to the number 15 on) are numbered). Then, this allows a detailed and univocal reconstruction of the evolution of the whole course in Theoretical Physics and, perhaps, it could be useful even for a further investigation on the last days of Majorana in Naples before his disappearance.
\\
Moreover, by analyzing even superficially the text contained in the Moreno Paper and by comparing it with that present in the Pisa manuscripts \cite{Bibliopolis}, it immediately comes out two important peculiarities: 1) the lectures contained both in the Moreno Paper and in the Pisa manuscripts are {\it completely} identical in the text, so that the copy performed by Moreno has to be considered as a faithful one; 2) the Moreno Paper contains {\it six} lectures not included in the documents of the Domus Galilaeana. For what concerns the scientific content and its style of presentation, this former conclusion is corroborated by a comprehensive analysis of the six previously unknown lectures based on the comparison with other writings by Majorana, mainly the five notebooks (``Volumetti'') \cite{volumetti} and the eighteen unpublished booklets (``Quaderni'') conserved in Pisa.

\begin{figure}
\begin{center}
\vspace{1.5truecm} \epsfysize=17.6cm \epsfxsize=12.5cm \epsffile{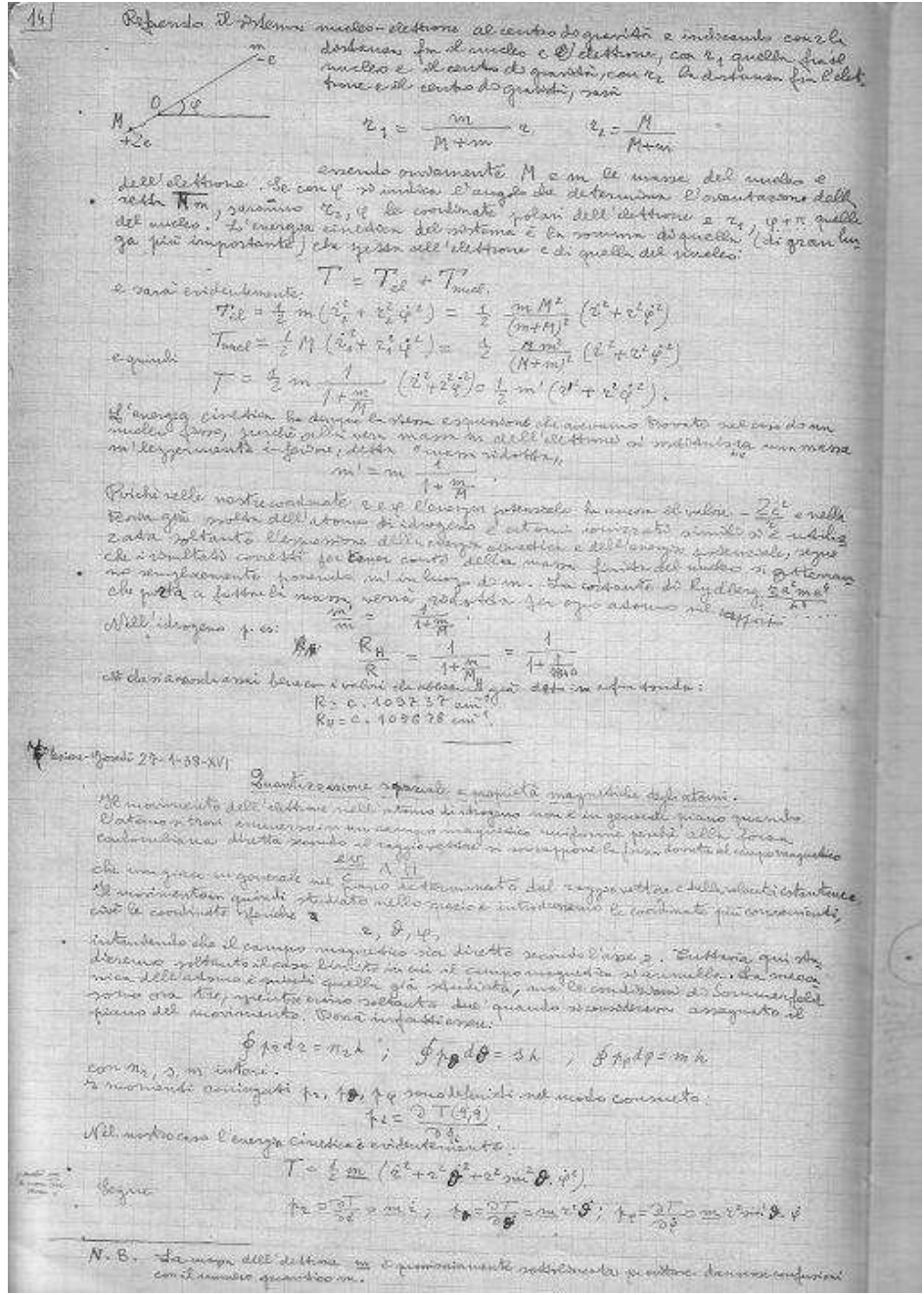}
\caption{One page of the Moreno Paper, where we find the beginning of the lecture N.7, the first previously unknown lecture by Majorana (the first part of the page appearing in it is, instead, present in the original manuscripts conserved in Pisa \cite{Bibliopolis}). Note the presence of the correction of a first writing of the lecture numbering and, for example, the addition of a marginal note below on the left.} \label{fig1}
\end{center}
\end{figure}

\subsection{From 1938 to now}

\noindent The lecture notes by Majorana have reached us by following different and not easily recognizable paths, so that some obscure points still remain.
\\
On Friday March 25 (the day before, Majorana held his 21st lesson), no lesson on Theoretical Physics was scheduled but, according to Senatore, ``contrarily to what he usually did, Majorana came to the Institute and remained there for few minutes'' \cite{Sena98}. It is likely that Majorana was aware of the habit of the students to convene to the Institute in order to study the topics treated at lesson the day before, so that he deliberately came there, that day of March, for providing his notes to the students \cite{Senatore}. ``From the corridor leading to the small room where I was writing, he called me by name: `Miss Senatore...'; he didn't entering in the room but remained in the corridor; I reached him and he gave me a closed folder telling: `that's some papers, some notes, take them... we will talk about later'; afterward he went away and, turned back, said again `we will talk about later' '' \cite{Sena98}. After that Friday March 25, Senatore did not go to the Institute of Physics (from a town near Salerno where she lived) for about 15 days due to health reasons, and the papers were conserved at her own home for a long time \cite{Senatore}. Senatore did not say anything on this affair for many months, but from another girl-student of the course (Minghetti) we now know that few days after the disappearance of Majorana, Carrelli asked the students for the notes of the course taken during the lessons \cite{Russo}. Maybe, Carrelli referred to the notes copied by the students from the originals, or even the notes taken directly from the students during the course; in any case it is remarkable the interest in such notes just few days after the disappearance of Majorana.
\\
When at the end of 1938 Senatore had close relations with Cennamo, the assistant to the Director Carrelli, the young girl-student showed the Majorana papers to Cennamo who, after some time and without saying anything to Senatore, brought them to Carrelli, probably for discussing some specific topic \cite{Senatore}. Carrelli, being the official depository of all Majorana assets in Naples, deemed appropriate to not return the papers to Cennamo.
\\
The subsequent story of the Majorana notes is quite obscure. From a letter dated February 2, 1964 from Gilberto Bernardini to Edoardo Amaldi (who asked him for some news about the notes) we learn as follows: ``These are the pages of the lecture notes by Majorana that I retrived, at long last, on my leaving from Geneva. They account for only a part. It seems to me that the remaining part was owned by Giovannino Gentile or, probably, was never wrote; more precisely, the confusing lines of a too far memory induce me to believe that Giovannino gave me this part for helping me in the understanding of the Dirac's textbook'' \cite{Bernardini}. Then, since 1965 Majorana's notes are owned by Amaldi who, in the following years, deposited them at the Domus Galilaeana together with some other manuscripts. No known document attesting the passage from Carrelli to Bernardini (or from Carrelli to Amaldi, assuming that Amaldi had the notes in his hands before Bernardini) exists. Nevertheless, it seems certain \cite{Senatore} that Senatore became acquainted again with the existence of the lecture notes by Majorana only after the death of her husband Cennamo, when Elio Tartaglione (one more former assistant to Carrelli), upon request of Cennamo still living, revealed to her how the notes passed on to Carrelli (as we reported above) and gave to her the just edited book with the lecture notes by Majorana \cite{Bibliopolis}. Thus, we can only make speculations on the passage of the Majorana papers from ``Naples'' to ``Rome'' which, however, will be not taken into account here. In any case, it should be noted that, if it is true that Bernardini received these notes from Gentile \cite{Bernardini}, they should be arrived at the Physics group in Rome before 1942, year of the premature death of Gentile (who was a sincere friend of Majorana, as well as the son of the famous Senator of the Italian Kingdom, well known to Carrelli).
\\
The path of the Moreno Paper is, in contrast, much more linear: the copy of the Majorana notes has been always kept by Moreno at home, and recently retrieved \cite{Moreno}.
\\
A subsequent detailed comparison between the two documents now available could perhaps put new light even on the missing part of the original manuscripts.

\subsection{Some features of the lectures in the Moreno Paper}

\noindent Moreno Paper's numbering and dating of each lecture held by Majorana in Naples between January and March of 1938, allow a sufficiently faithful reconstruction of the course on Theoretical Physics delivered by the great scientist. However, it should not be forgotten that only one source is available for many pieces of information, so that the discussion presented in the following could not be completely free from possible interpretational difficulties. In Table 1 we have reconstructed a table of contents of the Majorana lectures: number, date and argument of each lecture are reported as they appear in the Moreno Paper (except for the last row, corresponding to some material which is present only at the Domus Galilaeana). It is interesting to observe that Moreno made initially a mistake in the numbering of the lectures up to lesson N.16: a given written number was subsequently deleted and replaced with the following one. This can be easily explained by noting that Majorana numbered his lectures starting from the lesson following the opening one (i.e., the first lecture attended by the students), while in a first time Moreno considered the opening lecture as the first effective one. Such a ``mistake" is no longer present from the lecture N.17 on: by keeping in mind that the original manuscripts by Majorana report an explicit numbering only starting from the lecture N.15 (and, in this case, the Moreno Paper reports twice the numbering), we can conclude that the manuscripts corresponding to the lectures N.15 and N.16 (and, possibly, even the previous ones) came in the hands of Moreno not before the break in the academic year due to the Carnival holidays (see below), that is after March 8. In any case, it seems likely that Moreno did not acquire the whole set of lecture notes in only one occasion.
\\
As mentioned, the opening lecture was given on January 13, but the corresponding text is absent in both the Moreno Paper and Pisa manuscripts: it was found by Erasmo Recami in 1972 among the papers conserved by the Majorana family \cite{Recami}. In the Moreno Paper the note ``Introduction: topics to be covered in the course" appears, probably denoting the fact that Moreno was present to the opening lecture.
\\
The notes corresponding to the subsequent four lessons, held from Saturday January 15 to Saturday January 22, are not present in any known document. In correspondence to these lectures, the Moreno Paper has 3 blank pages, probably foreseeing to receive later the corresponding text. Their topics are unknown but, following Cabibbo \cite{Cabibbo}, we could reasonably assume that Majorana introduced the most relevant phenomenology of Atomic Physics, which then continued to be illustrated in the immediately following lectures. This first part of the course extended up to the break for the Carnival holidays, i.e. up to the lesson N.15 given on February 17.
\\
The text of the lectures ranging from N.6 to N.21 is entirely covered by the Moreno Paper; we address the interested reader to the discussion in \cite{Cabibbo} for an analysis of the content of the lectures which are present also in Pisa manuscripts.
\\
The lecture N.7 is the first one reported in the Moreno Paper: it is missing at the Domus Galilaeana. In this lecture Majorana studied the angular momentum quantization and introduced the concept of spin, which was the starting point for the following lectures on the periodic table of the elements and on the atomic spectra. The existence of such a lecture, not even suspected before the appearance of the Moreno Paper, confirms the attitude of Majorana to introduce in a very detailed and exhaustive way the novel concepts not yet studied by the students.
\\
The lesson scheduled for Tuesday February 1 was not held, probably for the concomitant 15th anniversary of the foundation of the Fascist Militia .
\\
The lecture N.10 held on February 5 is the second one whose notes are not present in Pisa manuscripts but are reported in the Moreno Paper. Cabibbo already speculated on the existence of such a lecture (or, at least, on the first part of it) \cite{Cabibbo}, since the preceding one (N.9) was evidently incomplete. In this lecture Majorana carried on the classical theory of radiation (a topic often considered by Majorana's personal studies \cite{volumetti}); in particular he discussed a peculiar (and simple) physical phenomenon - solar light scattering by Earth atmosphere - which hardly (and strangely) entered in a Theoretical Physics course. 
\\
The following four lessons, held from Tuesday February 8 to Tuesday February 15, again again are absent in the Domus Galilaeana collection; their existence was not suspected at all, since they cover an unexpected topic: the Theory of Special Relativity. This very important argument for Teoretical Physics still found some spare opposition in 1938 in Italy (just as an example, we point out the very critical attitude of a renowned physicist as Quirino Majorana, uncle of Ettore Majorana), and was not usually treated in Physics courses. From Moreno paper we thus learn that Majorana was the first University professor that introduced Relativity in a Physics course at the University of Naples \footnote{Note, however, that especially the General Relativity was covered in Mathematics courses as, for example, was done by the well known mathematician Roberto Marcolongo at the Institute of Mathematics of the University of Naples.}. In these four lectures, he treated it according to his custom, by starting from simple phenomenology and later introducing the corresponding mathematical formalism. 
\\
In the first Relativity lecture, Majorana first discussed Galilean Relativity, then through a discussion of Michelson \& Morley experiment passed to the question of the existence of aether, and at last introduced Lorentz transformations, with application to the electromagnetic field case. In the following lecture he then considered the formal developments of Einstein Relativity, and again came back to the electromagnetic field case, but now focusing on potentials rather than fields. Starting from the Fresnel formula for the Fizeau experiment on the measurement of the light speed, the problem of the relativistic sum of velocities is instead treated in the third lecture of the series, where the relativistic invariance of the electric charge is considered as well. Ranging from the end of the third Relativity lecture to the beginning of the fourth one, Majorana then passed to study (in a very detailed way) the relativistic dynamics of the electron, the starting point being a variational principle (see also \cite{volumetti}). The lectures missing in Pisa archive thus end with the discussion of both the photoelectric effect (and the Einstein interpretation of it) and of Thomson scattering. Just looking at this two last topics and at the following one in the subsequent lecture (the Compton effect), we can then suppose that Majorana introduced the Theory of Special Relativity in his course in order to give to the students a detailed and clear discussion of those mentioned topics that, although they belongs to Quantum Physics (the foremost argument of modern Theoretical Physics courses), did require an underlying use of the Special Relativity for a correct interpretation of them.
\\
The course break off on Thursday February 17 and restarted on Tuesday March 8 with the introduction of the mathematical formalism of Quantum Mechanics. Such a long break was partially due to the Carnival holidays (the corresponding academic holidays ranged from February 24 to March 2); the remaining days of vacation were, instead, probably related (at least in part) to some concomitant political and social (fascist) events, such as the visit in Naples of Bruno Mussolini (February 22) and the death of the famous poet Gabriele D'Annunzio (March 2).
\\
The text of the notes, corresponding to the lectures from N.15 to the last one N.21 and reported in the Moreno Paper, is identical to that of the original Pisa manuscripts; it has been already considered in \cite{Cabibbo} and here we do not indulge further on it. We only note, from an historical point of view, the postponement of the lesson scheduled for Tuesday March 15 ``for the visit of His Majesty the King" (as reported in the Moreno paper) \footnote{In this day the king Vittorio Emanuele III and the Ministry of Education Bottai inaugurated in Naples a renowned picture exhibition.} and that scheduled for Saturday March 19, cancelled for the celebrations of the ``St. Joseph day".
\\
The papers corresponding to Majorana lecture notes, conserved at the Domus Galilaeana, end with an unnumbered and undated (long) note, which is substantially different from the preceding ones (many deletions are present) and appear as a preliminary copy \cite{Bibliopolis}; it is not present in the Moreno paper. The course student Senatore interpreted it as the text prepared for the lecture scheduled for the day after his disappearance (see \cite{Sena98} as well as \cite{Cabibbo}). Such an hypothesis, although likely, is difficult to explain by looking at the topics studied extensively in this note (practically, Majorana dealt with some applications of Atomic and Molecular Physics), which seem unrelated to that considered in the previous lecture N.21.

\begin{figure}
\begin{center}
\vspace{1.5truecm}
\epsfysize=17.6cm \epsfxsize=12.5cm \epsffile{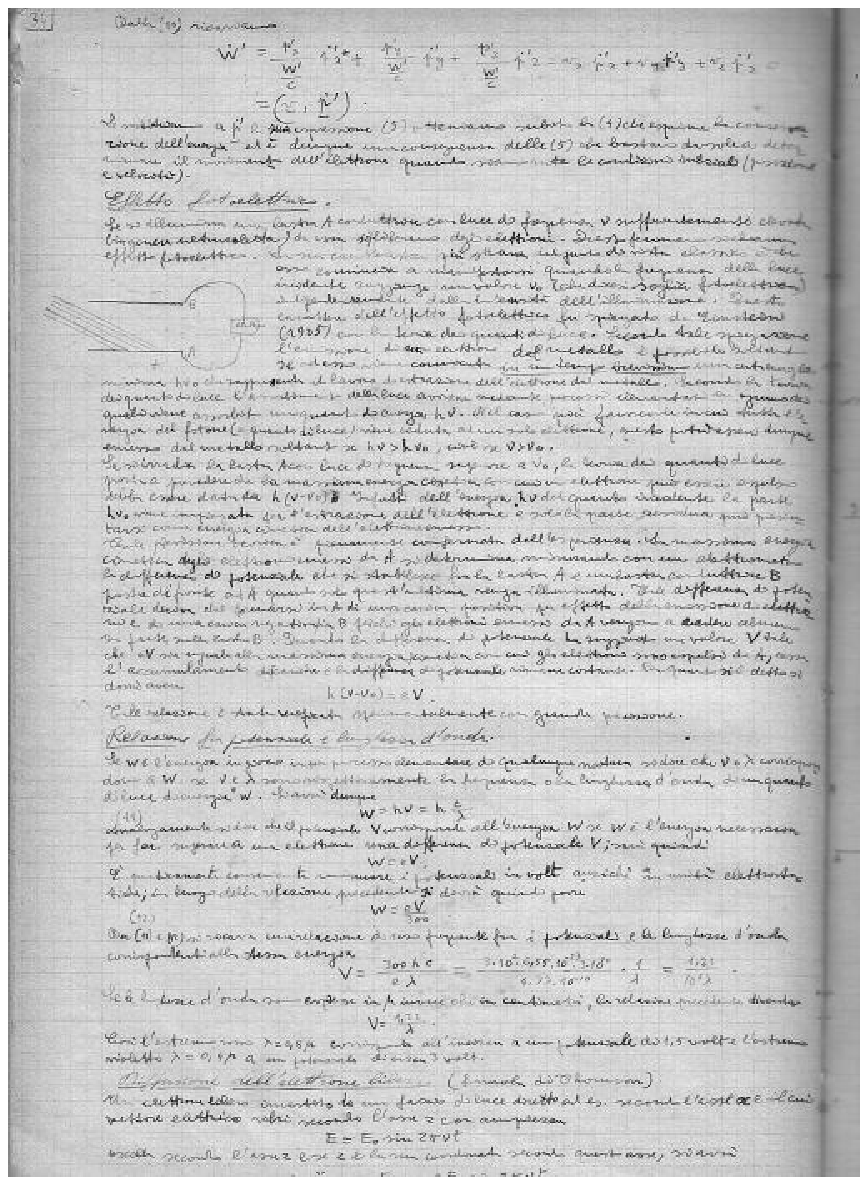}
\caption{Another page of the Moreno Paper, where Majorana ends the lecture N.14 (the last of the six lectures missing in the Pisa archive) with the application of the Theory of Special Relativity to the photoelectric effect. The following lecture N.15, which is also present in the Pisa original manuscripts \cite{Bibliopolis}, will continue with the study of the quantum effects whose explanation requires Special Relativity such as, for example, the Compton effect.}
\label{fig2}
\end{center}
\end{figure}

\section{Conclusions}

\noindent The discovery of the Moreno Paper, along with some memories allows a very detailed historical reconstruction of the Theoretical Physics course given by Ettore Majorana at the University of Naples.
\\
The relevance of such a recovering lies, first of all, in the fact that the 53 pages handwritten by Eugenio Moreno report, without doubt, the text of the lecture notes prepared by Majorana for his students. This conclusion clearly emerges from multiple comparisons of the Moreno Paper with the original manuscripts conserved at the Domus Galilaeana in Pisa, as well as with many other (some of them publicly available) works written by Majorana (mainly the five ``Volumetti" and the eighteen ``Quaderni"). From a strictly scientific point of view, the present discovery has led to the discovery of six previously unknown lectures, four of which deals with the Theory of Special Relativity, a completely unexpected topic for the Majorana Theoretical Physics course.
\\
In addition to the text copied from the original manuscripts, Moreno Paper shows also some remarkable notes, not present there and not arguable from other sources (if not in a partial and approximate way). They consist mainly \footnote{Also some marginal notes appear in the present document, written by Moreno in order to point out some improper notation or equation numbering, or even some apparently incomprehensible word written by Majorana. Such marginal notes are especially relevant for a detailed analysis of the text and contents of the Moreno Paper that, however, is not the aim of the present paper.} in the numbering and dating of each lecture, which has allowed to follow directly the development of the course, along with its own temporal deadlines, probably important even for a reconstruction of the last days of Majorana in Naples before his mysterious disappearance.
\\
The following steps will be mainly oriented to the detailed study of the scientific content of the six previously unknown lectures that, even at a preliminary analysis, appears as particularly interesting and promising.

\section*{Acknowledgments}

\noindent The tracking of the Moreno Paper has been possible through the agency of Cesare Moreno, to which our sincere gratitude is here addressed. We are also very grateful to E. Recami for his impressive photographic and documentary material made available to us, and to A. De Gregorio, F. Guerra, G. Longo, F. Lizzi, E. Majorana jr, G. Mangano, G. Miele, O. Pisanti and B. Preziosi for their kind collaboration.

\vspace{2truecm}


\begin{table}
\begin{center}
\begin{tabular}{|l|l|l|l|l|}
\hline \hline N. & Date $\hphantom{......}$ & M. P. & D. G. & Argument \\ \hline  
1. & \bmm Thursday \\ 13-1-38 \eem & missing & missing & \bbm
Opening lecture. \\ Introduction: topics to be covered in the course. \eem \\ \hline
2. & \bmm Saturday \\ 15-1-38 \eem & missing
& missing & \\ \hline
3. & \bmm Tuesday \\ 18-1-38 \eem &
missing & missing & \\ \hline
4. & \bmm Thursday \\ 20-1-38
\eem & missing & missing & \\ \hline
5. & \bmm Saturday \\ 22-1-38
\eem & missing & missing & \\ \hline
6. & \bmm Tuesday \\ 25-1-38 \eem & present & present &
\bbm - Fine structure formula. \\ Experimental verification and its validity. \\ Nucleus dragging. \eem \\
\hline
7. & \bmm Thursday \\ 27-1-38 \eem & \underline{present}
& missing &
\bbm - Spatial quantization and magnetic properties of atoms. \\ 
- Generalities on the alkali metal spectra and the spinning electron hypothesis. \eem \\
\hline
8. & \bmm Saturday \\ 29-1-38 \eem & present & present &
\bbm - Pauli or exclusion principle and the interpretation of the periodic table of the elements. \\
- Sommerfeld conditions for the computation of the energy levels of alkali metals. \eem \\
\hline
9. & \bmm Thursday \\ 3-2-38 \eem & present & present &
\bbm - The spectrum of atoms with two valence electrons.
\\ - Classical theory of radiation. \eem \\
\hline
10. & \bmm Saturday \\ 5-2-38 \eem & \underline{present} & missing & \bbm - Integration of the Maxwell equations with applications to the radiation of an oscillating system with small dimensions compared with the emitted wave-length. \\ - Solar light scattering by the atmosphere. \eem \\
\hline \hline
\end{tabular} \vspace{0.5truecm}
\end{center}
\caption{Reconstruction of a table of contents of the lectures given by Majorana at the University of Naples, as emerges from the Moreno Paper. In the first two columns the numbering and the date of a given lecture is reported, while in the subsequent two columns the presence of the text of that lecture in the Moreno Paper (M.P.) and/or in the manuscripts of the Domus Galilaeana (D.G.) is indicated. A line under the text points out the lectures which are non present in both the archives. Finally, in the last column, the title of the sections of each lecture in the M. P. (or in the manuscripts of the D.G.) is given.}
\end{table}

\setcounter{table}{0}

\begin{table}
\begin{center}
\begin{tabular}{|l|l|l|l|l|}
\hline \hline N. & Date $\hphantom{......}$ & M. P. & D. G. & Argument \\ \hline  
11. & \bmm Tuesday \\ 8-2-38 \eem & \underline{present} &
missing &
\bbm - The relativity principle in Classical Mechanics. \\
- Michelson and Morley experiment. \\
- Lorentz transformations. \eem \\
\hline
12. & \bmm Thursday \\ 10-2-38 \eem & \underline{present}
& missing &
\bbm - The relativity principle according to Einstein. \\
- Transformation laws for the electromagnetic potentials. \eem \\
\hline
13. & \bmm Saturday \\ 12-2-38 \eem & \underline{present} & missing & \bbm - Fresnel formula and Fizeau experiment. \\
- Electric charge invariance. \\
- Minkowski space. \\
- Equations of motion for an electron in an arbitrary electromagnetic field. \eem \\
\hline
14. & \bmm Tuesday \\ 15-2-38 \eem & \underline{present}
& missing &
\bbm - Relativistic dynamics of an electron. \\
- Photoelectric effect. \\
${}$ ~~~ - Relationship between the potential and the wave-length. \\
${}$ ~~~ - Free electron scattering (Thomson formula). \eem \\
\hline
15. & \bmm Thursday \\ 17-2-38 \eem & present & present
&
\bbm - Compton effect. \\
- Franck and Hertz experiment. \eem \\
\hline
16. & \bmm Tuesday \\ 8-3-38 \eem & present & present &
\bbm Notions on matrix algebra: \\
- Vector space in $n$ dimensions. \\
- Matrices and linear operators. \eem \\
\hline
17. & \bmm Thursday \\ 10-3-38 \eem & present & present
&
\bbm - Unitary systems. \\ - Hermitian operators. Hermitian forms. \eem \\
\hline
18. & \bmm Saturday \\ 12-3-38 \eem & present & present &
\bbm - Simultaneous diagonalization of commuting operators. \\
- Infinite matrices. \eem \\
\hline \hline
\end{tabular} \vspace{0.5truecm}
\end{center}
\caption{Continuation.}
\end{table}

\setcounter{table}{0}

\begin{table}
\begin{center}
\begin{tabular}{|l|l|l|l|l|}
\hline \hline N. & Date $\hphantom{......}$ & M. P. & D. G. & Argument \\ \hline  
19. & \bmm Thursday \\ 17-3-38 \eem & present & present &
\bbm - Fourier integrals. \\
Wave Mechanics. \\
- De Broglie waves. \eem \\
\hline
20. & \bmm Tuesday \\ 22-3-38 \eem & present & present
&
\bbm - Phase and group velocity. \\
- Non relativistic wave equation. Statistical interpretation of wave-packets. \eem \\
\hline
21. & \bmm Thursday \\ 24-3-38 \eem & present & present
&
\bbm - First generalization of the statistical interpretation and uncertainty relations. \eem \\
\hline
? & ? & missing & \underline{present} &
\bbm - On the meaning of quantum state. \\
- Symmetry properties of a system in Classical and Quantum Mechanics. \\
- Resonance forces between non-symmetrized states for small perturbations. Non combinable symmetry characters. \\
- Spectroscopic effects in atoms with two electrons. Resonance between equal potential wells and the theory of homeopolar valence according to the method of the binding electrons. \\
- Properties of symmetrized states which cannot be obtained from non symmetrized ones by means of a small perturbation. Alternating bands, hydrogen, ... .
 \eem \\
\hline \hline
\end{tabular} \vspace{0.5truecm}
\end{center}
\caption{Continuation.}
\end{table}

\end{document}